\documentclass[epj]{svjour}
\usepackage{graphicx}
\usepackage{amsmath,amssymb}

\begin{document}
\title{Dimer percolation and jamming on simple cubic lattice}
\author{Yu. Yu. Tarasevich\thanks{e-mail: \texttt{tarasevich@aspu.ru}} \and V. A. Cherkasova% etc
}                     % Do not remove
\institute{Astrakhan State University, 20a Tatishchev St., Astrakhan, 414056,
Russia}
\date{Received: date / Revised version: date}
\abstract{ We consider site percolation of dimers (``neadles'') on  simple
cubic lattice.  The percolation threshold is estimated as  $p_c^\text{perc}
\approx 0.2555 \pm 0.0001$. The jamming threshold is estimated as
$p_c^\text{jamm} = 0.799 \pm 0.002$.
 % end of PACS codes
} %end of abstract
\maketitle
\section{Introduction}\label{intro}

Percolation theory deals with forming of connected objects inside disordered
media. One of the possible kinds of percolation problems and at the same time
more often used and simplest is site percolation. In general, site percolation
is defined on a lattice (graph) in $d$-dimensional space where each site (node)
can be either occupied with the probability $p$ or empty with the probability
$1-p$. Neighboring occupied sites form a cluster. If a cluster is so large that
it reaches the two opposite edges of the lattice, the cluster is called
percolating (spanning or connecting). The lowest concentration of occupied
sites for which there is a percolating cluster is called the percolation
threshold $p_c$ for a particular lattice~\cite{Stauffer}.

Percolation is a critical phenomenon. It is a purely geometric phase transition
closely connected with usual second-order phase transition. Percolation theory
is very simple but general, powerful and useful tool. It attracts attention of
researchers (mathematicians, physicists, programmers, engineers) because of
different reasons from pure theoretical to applied ones. Percolation theory has
been successfully applied to a wide number of problems in a large variety of
fields~\cite{Stauffer,Sahimi,Grimmet,Kesten}. One of such applications is phase
transition from sol to gel (see e.g.~\cite{Coniglio79,Rottereau2003}). There
are different modifications of percolation problems used to describe sol-gel
phase transition. Usually occupied sites represent monomers and empty sites are
associated with solvent molecules.

Most of the studies are devoted to the random (Berno\-ulli) percolation of
particles (sites) with single occupancy. Nevertheless, the percolation of
$k$-mers has been intensively studied during last decade. The percolation of
$k$-mers may be described as a kind of correlated percolation when particles
occupy several ($k$) contiguous lattice sites~\cite{Dolz2005EPJB}. Recently,
different kinds of $k$-mers percolation problems on square lattice have been
investigated~\cite{Dolz2005EPJB,Dolz2005PRE,Cornette,Vandewalle,Kondrat2001a,Kondrat2001b}.

Another realization of the percolation problem is random sequential adsorption
(RSA) in which objects (point particles, segments, rectangles, needles, etc.)
are put on randomly chosen sites and the objects do not move. It is also
possible to consider RSA in a continuum \cite{Evans1993,Nakamura,Loscar}.

In filling process, objects of finite size are randomly deposited on an
initially empty substrate or lattice with the restriction that they must not
overlap with previously added objects. Due to the blocking of the lattice by
the already randomly adsorbed elements, the limiting or "jamming coverage" is
less than that corresponding to the close packing. More recently, leading
contributions have been presented
in~\cite{Vandewalle,Kondrat2001a,Kondrat2001b,Rampf2002} treating with the
relationship between the jamming coverage and the percolation threshold. In
particularly, Vandewalle et al. \cite{Vandewalle} have found for the ``needle''
that the ratio of the two threshold concentrations $p^{perc}_c$ and $p^{jam}_c$
is constant regardless of the length of the needle
$$p^{perc}_c/p^{jam}_c = 0.62 \pm 0.01.$$

In the present paper we extend the study of dimer percolation and jamming to
simple cubic lattice lattices in the framework of a MC analysis. A study of the
finite size effects is presented. The main aim of the paper is to determine the
percolation threshold.

\section{Numerical results: simulation scheme, estimation of percolation probabilities and
finite-size scaling analysis} \label{sec:results}

The numerical results were obtained using Hoshen--Kopel\-man
algorithm~\cite{Hoshen76}. We investigated a number of sample lattices with
linear size $L$ up to 128 sites. Free and mixed boundary conditions were
utilized. In the case of mixed boundary condition, periodic boundary conditions
were applied along two directions, other boundaries are suppose to be open. We
were looking for a cluster spanning two opposite free edges.

We used long period ($> 2\cdot 10^{18}$) random number generator of
L'Ecuyer~\cite{Press} to fill a lattice with the dimers. Sample lattice was
swept site by site. We try to fill the empty sites with the given probability
$p$ by randomly orientated dimers. Not all attempts to place a dimer are
successfully, indeed. When the lattice is filled, we check the actual part of
the filled sites. If this quantity is smaller than the given probability $p$,
the filling process is started once again. There is no possibility to fill a
lattice with the necessary probability if $p$ is large enough. We can associate
the highest possible actual part of the filled sites  with jamming threshold
$p_c^\text{jamm}$.

\begin{figure}[!htbp]
\centering
\includegraphics*[width=\linewidth]{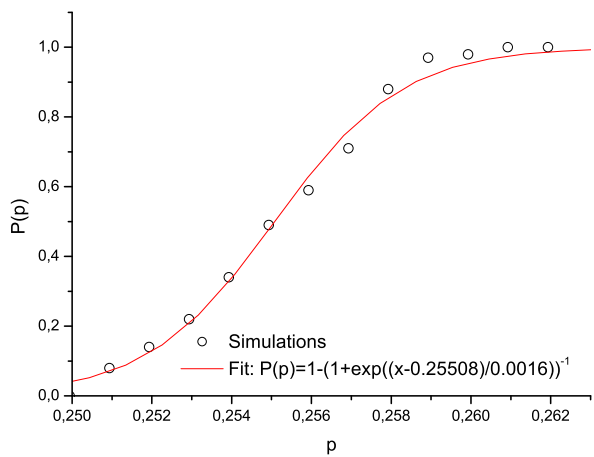}\caption {
Probability to find percolating cluster. Linear size of the lattice
$L = 128$. Mixed boundary conditions.}\label{fig:threshold}
\end{figure}

Estimates for percolation threshold $p_c$ have been obtained by means of
percolation frequencies. Simulations give the percolation frequencies $P(p)$,
which serve as an approximation of the percolation probability. Critical
percolation have been estimated by nonlinear fit functions defined by
\begin{equation}
P(p) =1 - \left(1 + \exp \left(\frac{p - p_c}{a}\right)\right)^{-1}.
\label{eq:fit1}
\end{equation}
This function reduces to a step function, if $a \to 0$. Percolation frequency
$P(p)$ for particular lattice of linear  size $L = 128$ and mixed boundary
conditions (periodic along two directions and free along one direction) is
shown with high resolution in Fig.~\ref{fig:threshold}.

\begin{figure}[!htbp]
\centering\includegraphics*[width=\linewidth]{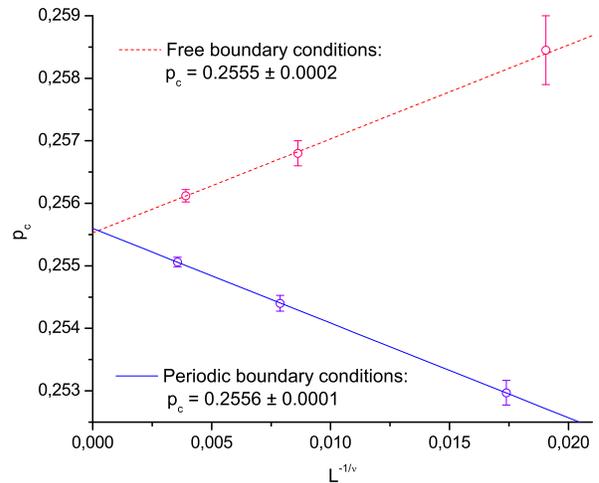}\caption
{Determination of the percolation threshold in the thermodynamical limit ($L
\to \infty$) using the scaling relation \eqref{eq:scal}}\label{fig:scaling}
\end{figure}

The percolation threshold $p_c(L)$ was calculated for three values of the
linear lattice size $L = 32, 64, 128$. The percolation threshold $p_c(\infty)$
for infinite lattice can be found by fitting these results for different
lattice sizes to the scaling relation
\begin{equation}
   \left| p_c(L) - p_c(\infty) \right|  \propto  L^{-1/\nu},
   \label{eq:scal}
\end{equation} where the
critical exponent $\nu$ has the value 0.875 in three dimensions~\cite{Bunde}.
This method leads to an estimate $p_c(\infty) \approx 0.2555$
(Fig.~\ref{fig:scaling}). The results obtained for two different kind of
boundary conditions are equal within the error bar.

\begin{figure}[!htbp]
\centering
\includegraphics*[width=\linewidth]{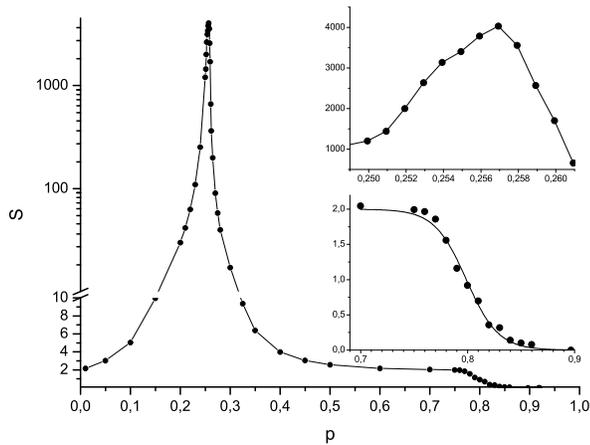}\caption {
Average cluster size. Linear size of the lattice $L = 128$. Mixed
boundary conditions.}\label{fig:s}
\end{figure}

Average cluster size as a function of $p$ is shown in Fig.~\ref{fig:s}. It
demonstrates typical behavior near the percolation threshold. If the
concentration of the dimers vanishes, the average cluster size goes to 2, i.e.
there are only isolated dimers. If the concentration is much greater than $p_c$
but smaller than jamming, the average cluster size tends to 2 again. It means
that there are the percolating cluster and very rare isolated dimers in the
holes inside the cluster.

We suppose that smooth threshold near the value $p = 0.8$ is finite size
effect. This threshold has to be extremely sharp if $L \to \infty$ and
corresponds to jamming. Our estimation gives $p_c^\text{jamm} = 0.799 \pm
0.002$. It means, that in contrast with percolation and jamming on square
lattice, the ratio
$$p_c^\text{perc}/p_c^\text{jamm} \approx 0.32.$$

\begin{figure}[!htbp]
\centering
\includegraphics*[width=\linewidth]{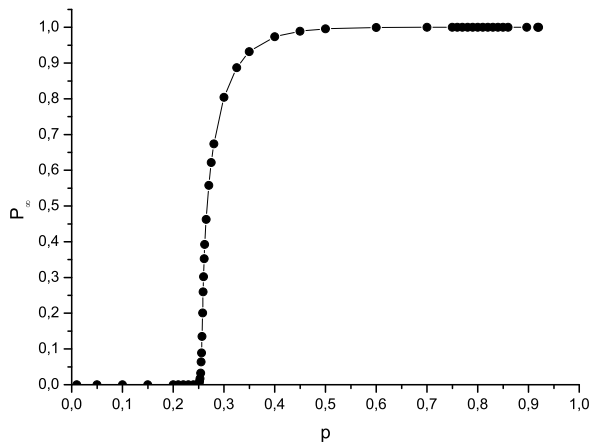}\caption {
Probability that an occupied site belongs to the percolating
cluster.}\label{fig:Pinf}
\end{figure}

Probability $P_\infty$ that an occupied site belongs to percolating cluster
goes to 1 very rapid above $p_c$ (Fig.~\ref{fig:Pinf}). It means that if one
tries to add another dimer in the system above the percolation threshold, the
dimer attaches with high probability to the existing percolating cluster.

\section{Discussion}

We investigated here new percolation problem. Except pure theoretical interest,
this problem may be useful to describe sol-to-gel phase transitions. In many
cases description of solute molecules as the point objects  is too
pared-down~\cite{Garboczi}. Consideration of the molecules as dimers
(``needles'') looks like more realistic in some situations. We hope that new
percolation model will serve for better description of sol-gel transitions.

One of the possible system for application of the model is desiccated aqueous
solution of albumen. The albumen molecules have rather complex shape.
Conventional site percolation is too pore model in this case. Dimer can be used
as the next approximation in comparison with point mole\-cu\-les. When water
evaporates, the concentration of albumin increases and phase transition from
sol to gel arises. Percolation threshold obtained in our work gives estimation
of the critical concentration. Jamming thresholds may be considered as
estimation of maximal possible part of solids in the gel matrix.

\section{Acknowledgments}
The authors are grateful to the Russian Foundation for Basic Research for
funding this work under Grant No.~06-02-16027-a.

\end{document}